# Defective graphene as promising anode material for Na-ion battery and Ca-ion battery

Dibakar Datta[1], Junwen Li[2], Vivek B. Shenoy[2,3,*]

[1] School of Engineering, Brown University, Providence 02912, USA
[2] Department of Materials Science and Engineering, University of Pennsylvania, Philadelphia PA 19104, USA
[3] Department of Mechanical Engineering and Applied Mechanics, University of Pennsylvania, Philadelphia PA 19104, USA

[*] Corresponding author: Vivek B. Shenoy (vshenoy@seas.upenn.edu)

## Abstract

We have investigated adsorption of Na and Ca on graphene with divacancy (DV) and Stone-Wales (SW) defect. Our results show that adsorption is not possible on pristine graphene. However, their adsorption on defective sheet is energetically favorable. The enhanced adsorption can be attributed to the increased charge transfer between adatoms and underlying defective sheet. With the increase in defect density until certain possible limit, maximum percentage of adsorption also increases giving higher battery capacity. For maximum possible DV defect, we can achieve maximum capacity of 1459 mAh/g for Na-ion batteries (NIBs) and 2900 mAh/g for Ca-ion batteries (CIBs). For graphene full of SW defect, we find the maximum capacity of NIBs and CIBs is around 1071 mAh/g and 2142 mAh/g respectively. Our results will help create better anode materials with much higher capacity and better cycling performance for NIBs and CIBs.

**Introduction**

Large-scale energy storage is of utmost importance for human advancement. For this purpose, rechargeable lithium ion batteries (LIBs) have been extensively used in portable electronics, light vehicles, and miscellaneous power devices over the last decade[1]. In terms of energy density, the seemingly ubiquitous LIBs exhibit superb performance as compared to other types of rechargeable batteries[2-5]. However, among light metals, Li is a very rare element. Its concentration in the upper continental crust is estimated to be 35 ppm[6]. Hence in recent time, there have been great concerns that available Li resources buried in the earth would not be sufficient to meet the ever increasing demands for LIBs[7]. These concerns have led to the active search for suitable alternatives[8]. Among these, sodium ion batteries (NIBs)[8, 9] and calcium ion batteries (CIBs)[10, 11] have drawn wide attention in recent years.

Though the energy density of an NIB is generally lower than a LIB[7], high energy density becomes less critical for battery applications in large-scale storage[9]. More importantly, the abundance and low cost of Na in the earth (10,320 ppm in seawater and 28,300 ppm in the lithosphere)[12, 13] and low reduction potential (-2.71V vs. Standard Hydrogen Electrode (SHE)) provide a lucrative low-cost, safe, and environmentally benign alternative to Li in batteries[14-16]. Like NIBs, CIBs offer several benefits such as low cost, natural abundance, chemical safety, low reduction potential (-2.87 V vs. SHE) and lighter mass-to-charge ratio[11, 17]. The use of polyvalent cations is the key to obtaining much larger discharge capacities than those of LIBs[10]. Moreover, nature stores energy with Na, Ca ions, not Li ions[18].

Electrochemical properties of the electrode materials are the cynosure of important battery performance characteristics such as specific capacity and operating voltage[9]. Hence the major challenge in advancing NIB and CIB technologies lies in finding better electrode materials. The best starting point is the investigation of the structure and chemistries of electrode materials that function well for Li intercalation. Graphite, the most widely used anode material for LIBs, has relatively low gravimetric capacity. Even for NIBs and CIBs, use of graphite yields very low capacity[19]. Recent experimental studies show that if we can lower the dimensionality of the conventional anode materials via nanotechnology, we can achieve higher capacity. For example, low dimensional materials like graphene[20, 21] and its oxide[22], carbon nanotubes[23, 24], and silicon nanowires[25] have been widely investigated as a possible replacement for graphite in LIBs.

Among the low-dimensional materials, graphene has attracted enormous attention ever since its discovery in 2004[26]. Besides its fascinating physical properties, it also shows considerable promise as atom/molecule containers for the potential applications on electrochemical storage devices[27-29]. However, impurities and defects, both stone-wale (SW) and divacancy (DV), are always present in graphene[30-32]. Recent studies discovered several structural defects in graphene at atomic resolution using the transmission electron microscope (TEM)[33, 34] and the scanning tunneling microscope (STM)[35, 36]. Structural defects have a strong influence on the electronic, optical, thermal, and mechanical properties of graphene[30]. A recent DFT study[37] predicted that the presence of defects would enhance the Li adsorption on graphene giving higher gravimetric capacity. Hence we have an open question: how will defects in graphene influence the adsorption of Na and Ca? To answer this question in detail, we have carried out the first-principles calculations based on DFT to thoroughly investigate the Na and Ca adsorption on graphene with various percentages of DV and SW defects.

## Methodology

All calculations are performed using the Vienna Ab Initio Simulation Package (VASP) [38] with the Projector Augmented Wave (PAW)[39, 40] method and the Perdew-Burke-Ernzerhof (PBE)[41] form of the generalized gradient approximation (GGA) for exchange and correlation functional. An energy cutoff of 600 eV was used in the plane wave expansion of wave functions. The Brillouin zones of 4×4 and 5×5 super cell are sampled with the Γ-centered k-point grid of 9×9×1 and 7×7×1 respectively. In order to avoid the spurious coupling effect between periodic graphene layers along the normal direction, the vacuum separation in the model structure is set to 18 Å. All atoms and cell-vectors are relaxed with a force tolerance of 0.02eV/Å.

The potential $V$ is defined as[42]

$$V = -\frac{\Delta G_f}{z \bullet F} \quad (1)$$

where the change in Gibb's free energy is

$$\Delta G_f = \Delta E_f + P\Delta V_f - T\Delta S_f \quad (2)$$

Since the term $P\Delta V_f$ is of the order of $10^{-5}$ eV[42], whereas the term $T\Delta S_f$ is of the order of the thermal energy (26 meV at room temperature), the entropy and the pressure terms can be neglected and the free energy will be approximately equal to the formation energy $\Delta E_f$ obtained from DFT calculations. The formation energy is defined as

$$\Delta E_f = E_{X_nG} - (nE_X + E_G) \quad (3)$$

Where $n$ is the number of X (X = Na, Ca) atoms inserted in the computational cell, $E_{X_nG}$ is the total energy of the Na/Ca intercalated graphene, $E_X$ is the total energy of a single Na/Ca atom in elemental BCC Na/ FCC Ca and $E_G$ is the total energy of a particular graphene structure. We have computed equilibrium energy for Na and Ca as -1.307 eV and -1.980 eV. If the energies are expressed in electron volts, the potential of the $Na_nG$ structures vs. $Na/Na^+$ as a function of Na content (and $Ca_nG$ vs. $Ca/Ca^{2+}$ as a function of Ca content) can be obtained as[42]

$$V = -\frac{\Delta E}{n} \quad (4)$$

The composition range over which Na/Ca can be reversibly intercalated determines the battery capacity.

## Results and Discussion

First we discuss the defective graphene systems we investigated for sodiation and calciation. Single vacancies (SV) with a carbon atom missing in graphene (or in the outermost layer of graphite) have been experimentally observed using TEM[33, 43] and STM[35]. However, Meyer *et al.*[33] showed that SV undergoes a Jahn-Teller distortion, which leads to the saturation of two of the three dangling bonds toward the missing atom.

For reasons of geometry, one dangling bond always remains. The SV appears as a protrusion in STM images due to an increase in the local density of states at the Fermi energy, which is spatially localized on the dangling bonds[35]. It is intuitively clear that the formation energy of such a defect is high because of the presence of an under-coordinated carbon atom. Hence instead of SV defects, we have concentrated on DV defects, where no dangling bond is present. The atomic network remains coherent with minor perturbations in the bond lengths around the defect. Simulations[44, 45] indicate that the formation energy $E_f$ of a DV is of the same order as for an SV (about 8 eV). As two atoms are now missing, the energy per missing atom (4 eV per atom) is much lower than for an SV. Hence, a DV is thermodynamically favored over an SV. Moreover, DV defects are the most common type of vacancy defects observed experimentally[46, 47] and as mentioned before, structures with any other kind of vacancy defect with dangling bonds are not stable[48]. As shown in Fig.1, DV defect can be obtained by removing C-C dimers from pristine graphene.

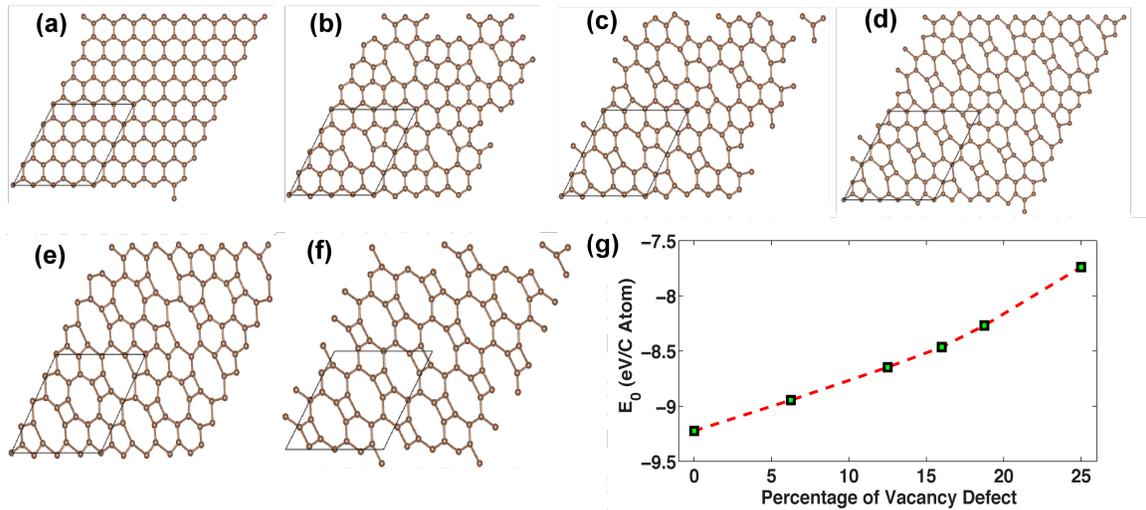

**Fig 1. (a) Pristine graphene and graphene with DV defects: (b) 6.25%, (c) 12.50%, (d) 16.00%, (e) 18.75% and (f) 25%. Systems shown here are of 2X2 size with periodicity in in-plane dimensions. Super cell used in the calculation is marked in black. All systems are relaxed structure. (g) Equilibrium energy per carbon atom for different percentages of DV defect.**

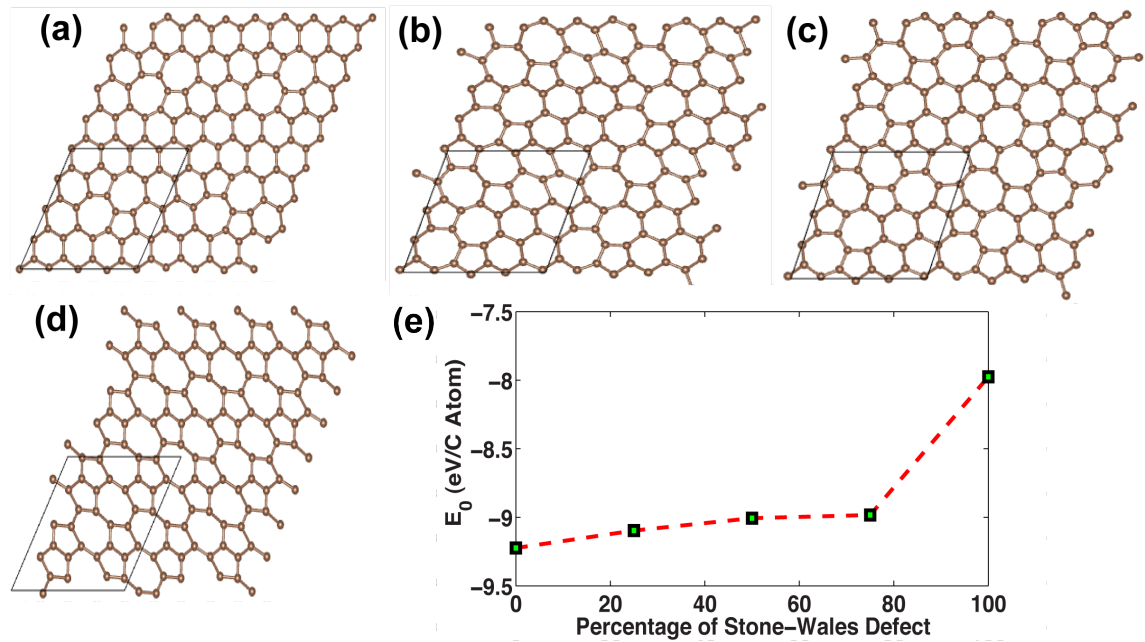

Fig 2. Graphene with SW defects: (a) 25%, (b) 50%, (c) 75%, (d)100%. Systems shown here are of 2X2 size with periodicity in in-plane dimensions. Super cell used in the calculation is marked in black. All systems are relaxed structure. (e) Equilibrium energy per carbon atom for different percentages of SW defect.

Five different percentages of defects are considered here: 6.25% (Fig. 1b), 12.50% (Fig. 1c), 16.00% (Fig. 1d), 18.75% (Fig. 1e) and 25% (Fig. 1f). All the systems shown here are relaxed structures. Fig.1g shows that equilibrium energy per carbon atom gradually decreases with the increase in DV defects.

Like DV, SW defects are another common type of structural defects observed experimentally[49]. The SW (55-77) defect has a formation energy $E_f$ = 5 eV[49]. The defective structure retains the same number of atoms as pristine graphene, and no dangling bonds are introduced. As shown in Fig.2, we have considered four types of SW defects with different defect concentration: 25% (Fig.2a), 50% (Fig.2b), 75% (Fig.2c) and 100% (Fig.2d). For 100% SW defect, we have the Haeckelite structure[50], which is a sheet full of 5-7 rings. The equilibrium energy per carbon atom is much less in this configuration (Fig. 2e).

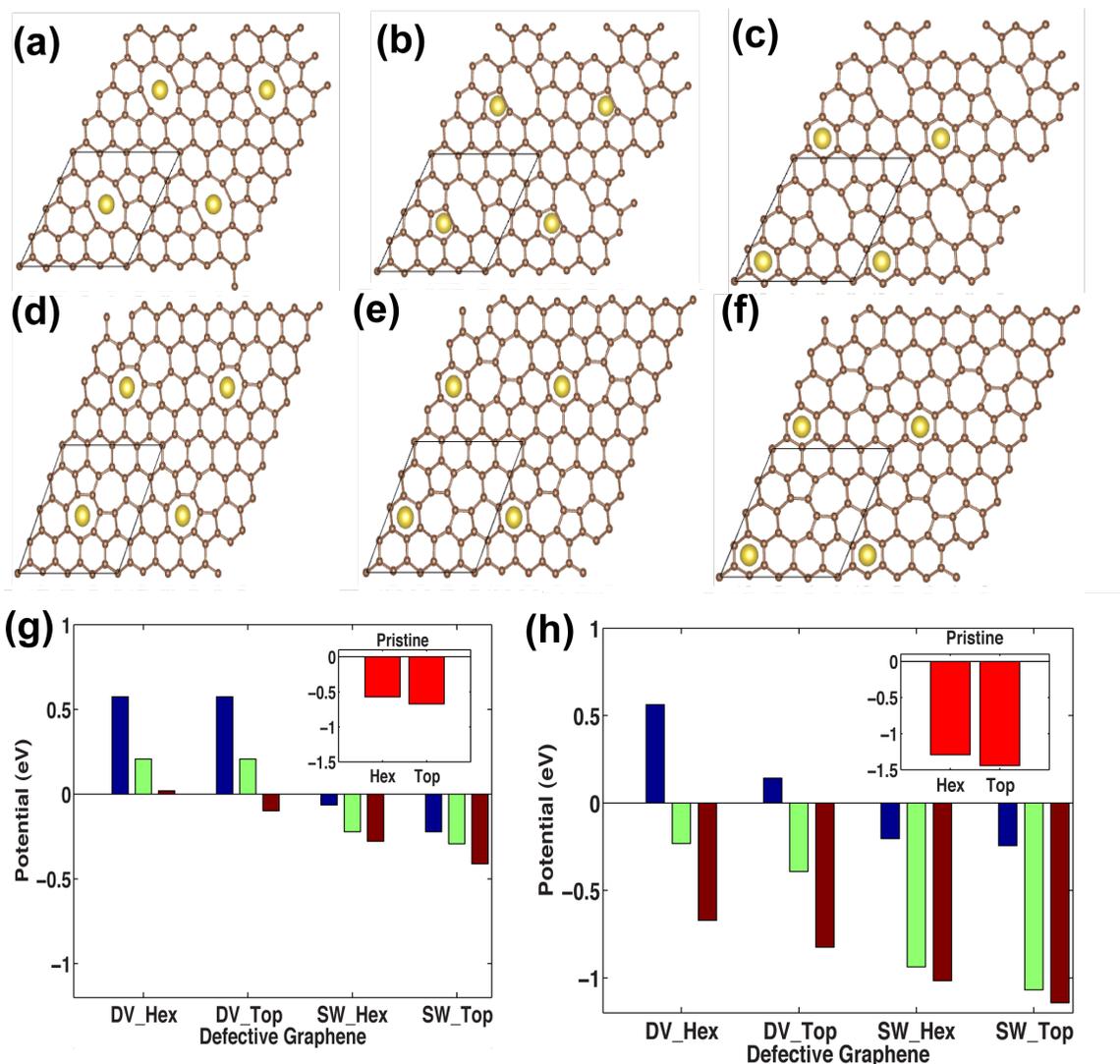

Fig 3. Na adsorption on graphene with (a-c) 6.25% DV defect (d-f) 25% SW defect: adatom (a,d) over the defect (O-position), (b,e) neighborhood of defect (N-position), (c,f) away from defect (A-position). (g) Sodiation & (h) Calciation potential for Na/Ca adsorption on different locations: Pristine graphene (inset) and graphene with DV and SW defects on Hex and Top sites. For each sites, three positions: O (blue), N (green) and A (brown) are shown.

**Table 1: Sodiation and Calciation potential (V in eV) for different positions of adatom at different sites in defective graphene.**

| Defect | | DV | | | | | | SW | | | | |
|---|---|---|---|---|---|---|---|---|---|---|---|---|
| Site | | Hex | | | Top | | | Hex | | | Top | |
| Positions | | O | N | A | O | N | A | O | N | A | O | N | A |
| V in eV | Na | 0.574 | 0.207 | 0.019 | 0.574 | 0.207 | -.098 | -.063 | -0.221 | -0.278 | -.221 | -.292 | -.410 |
| | Ca | 0.562 | -0.231 | -0.671 | 0.142 | -.391 | -.824 | -.204 | -0.937 | -1.016 | -.243 | -1.069 | -1.14 |

We first focus on pristine and lowest defect density. The lattice constant of graphene is 2.46 Å[51, 52]. We consider two sites of high symmetry for adsorption: the site on the top of a carbon atom (Top) and the site at the center of a hexagon (Hex) of graphene sheet. The inset in Fig. 3g and Fig. 3h (in red) shows the sodiation and calciation potential for pristine graphene respectively. The negative potential indicates that adsorption is not possible. Next we investigate the influence of lowest defect density: 6.25% DV defect and 25% SW defect. For both Hex and Top sites, we consider three positions: *over the defect (O-position), neighborhood of the defect (N-position), and away from the defect (A-position)*. Figs.3a-3c & Figs.3d-3f show Na on O (Fig. 3a, 3d), N (Fig. 3b, 3e), and A (Fig. 3c, 3f) position at the Hex site of graphene with DV defect & SW defect respectively. Similarly, we consider O, N and A positions of Top site.

The sodiation and calciation potentials for three different positions (O, N, A) for both Hex and Top sites are summarized in Table 1. The information in the Table is condensed in Fig 3g (sodiation potential) and Fig. 3h (calciation potential). For *DV_Hex* (Hex site of DV defect) and *DV_Top* (Top site of DV defect), we notice that O position (blue), as expected, is the most favorable position for adsorption. Sodiation potential is reduced to zero or negative from O to A position. For SW defect, lowest defect density (25%) does not favor Na adsorption for any location. However, the O-position has less negative potential compared to N & A positions. The same procedure applies for the calculations for Ca and similar trend is obtained as shown in Fig. 3h. It is clear that adatoms tend to cluster around the defective zone.

In order to obtain insight on the adsorption on defective sheets, we perform bonding charge density analysis[53]. Fig.4 shows the bonding charge density passing through the bond between Na/Ca and the nearest carbon atom. The bonding charge density is obtained as the difference between the valence charge density of strain-free graphene-Na/Ca sheet and the superposition of the valence charge density of the constituent atoms. A positive value (red) indicates electron accumulation while a negative value (blue) denotes electron depletion. These changes in bonding charge distributions after introduction of defects clearly show the enhanced charge transfer from Na/Ca to graphene sheet leads to adsorption of adatoms.

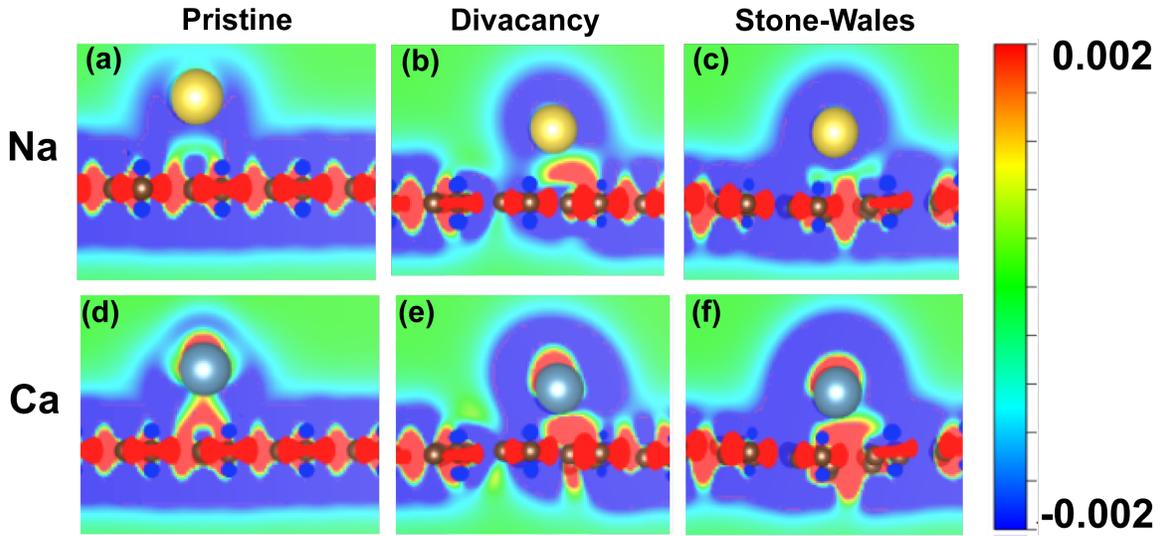

**Fig 4. Bonding charge density for Na & Ca (*DV_Top* site & O-position) for Pristine (a,d), Divacancy (b,e) and Stone-Wales (c,f) system obtained as the charge density difference between the valence charge density before and after the bonding. Red and blue colors indicate the electron accumulation and depletion, respectively. The color scale is in the units of e/Bohr$^3$.**

**Table 2: Charge transfer from Na/Ca to graphene.**

| Ion | Pristine | Divacancy | Stone-Wales |
|---|---|---|---|
| $Na^+$ | 0.6617 | 0.8848 | 0.8073 |
| $Ca^{+2}$ | 0.8208 | 1.3727 | 1.1189 |

The charge redistribution can be quantitatively estimated by computing the charge transfer using Bader charge analysis. Table 2 shows the magnitude of the charge transfer for different positions. In case of $Na^+$ ion, charge transfer to pristine graphene is 0.6617e while for structure with DV & SW defect; the transferred charges are increased to 0.8848e and 0.8073e respectively. For $Ca^{+2}$, corresponding charge transfer is 0.8208e, 1.3727e and 1.1189e respectively. For each case, DV defect case has more charge transfer resulting in more adsorption of adatoms.

From our results explained in Fig. 3 and Fig. 4, we have discovered that the O-position of Hex-site is the most favorable location of adsorption. Hence, we primarily focus on this location while initially distributing the Na/Ca adatoms. Still for every case, there are many possibilities of initial distribution. For each case, we have considered three different initial configurations to get the potential range and reported the average value. It is obvious that at low concentration, more possibility of initial distribution leads to wider range of potential. For each percentage of defects, we have carried out DFT calculations for different Na/Ca concentration until we cross the maximum limit of capacity i.e. when potential becomes -ve.

Fig.5a summarizes the sodiation potential for five different DV defect percentages. For each defect density, potential is dropped with the increase of Na concentration. For higher defect density, potential is more for a given Na concentration and the maximum percentage of adsorbed Na is increased. As shown in Fig. 1f, 25% is the maximum DV defect density possible. Beyond this limit, structure will have dangling bond[48]. Fig. 5b shows one of the configurations of $Na_8C_{26}$ where Na adatoms are mainly located on and around O-positions i.e. adatoms tend to cluster around the defective zone. As shown in Fig. 5c, for SW defect, percentage of adsorption is increased with the increase in defect density. Fig 5d shows one of the configurations of $Na_6C_{32}$. Results for Calcium adsorption are summarized in Fig. 6. We notice that qualitatively the adsorption behavior of Ca in DV and SW graphene is same as Na.

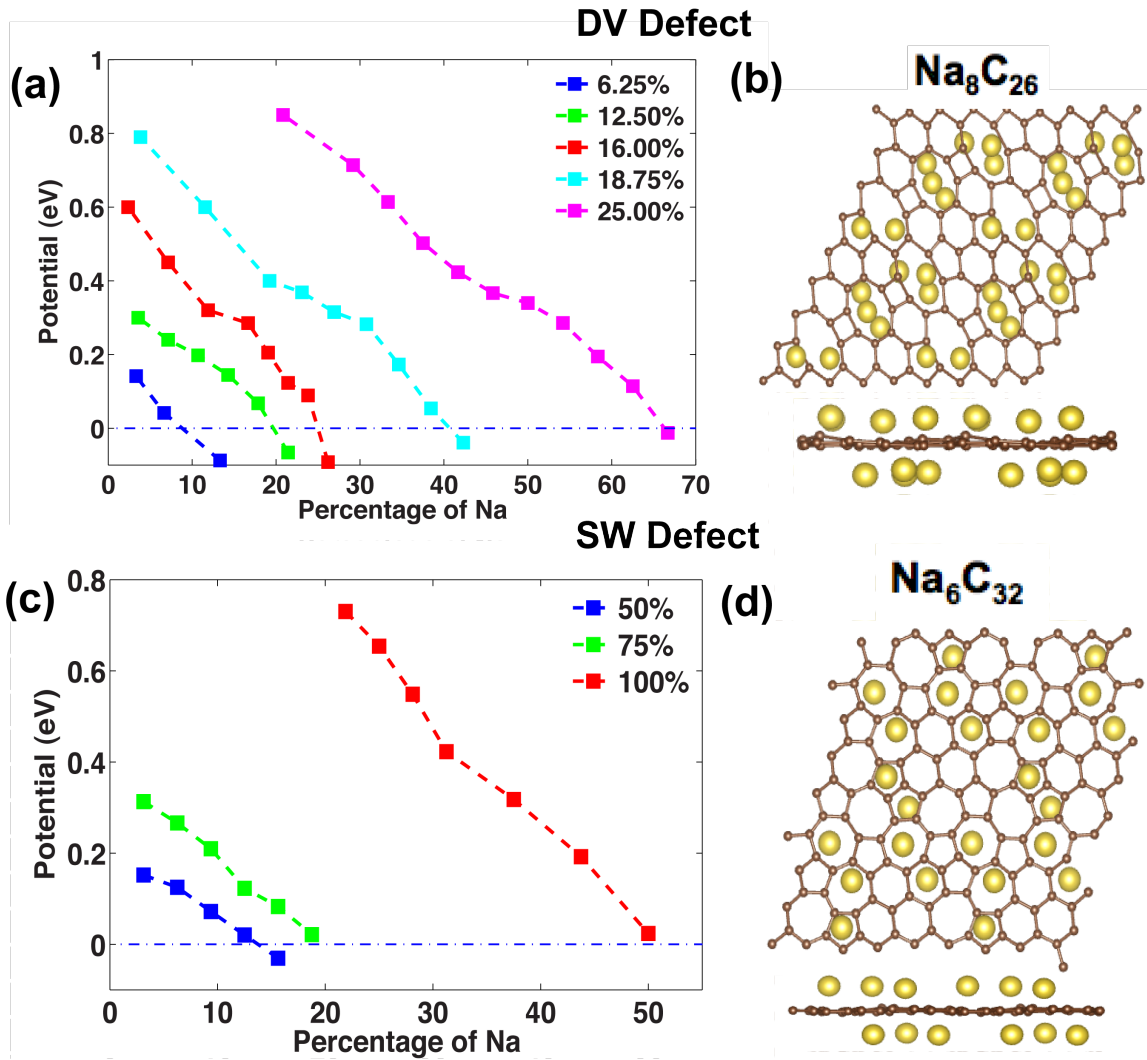

**Fig 5. Sodiation potential for different percentage of Na adsorbed for different percentages of (a) DV and (c) SW defect. Top and Side view of one of the (b) $Na_8C_{26}$ and (d) $Na_6C_{32}$ relaxed configurations.**

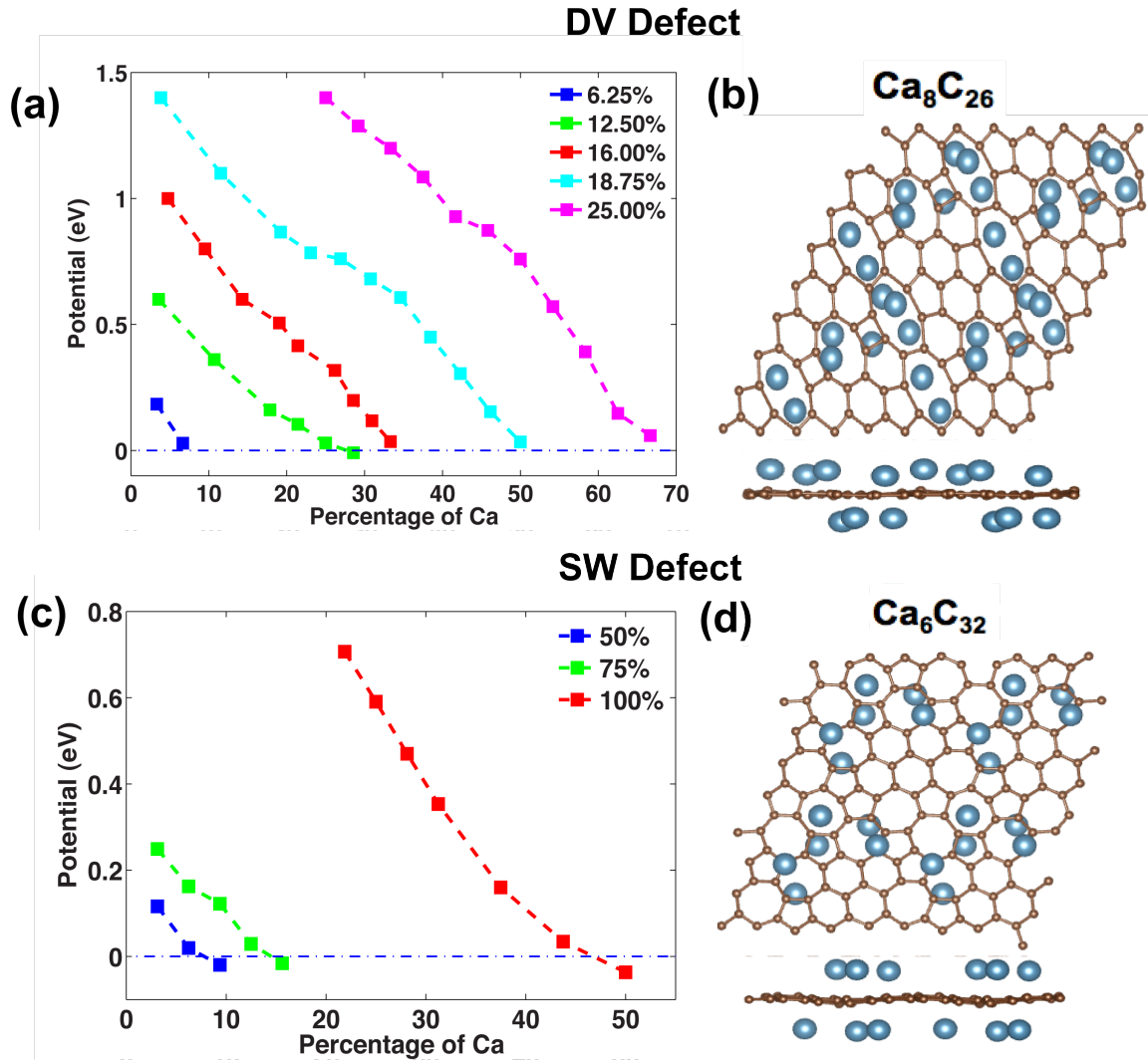

Fig 6. Calciation potential for different percentage of Ca adsorbed for different percentages of (a) DV and (c) SW defect. Top and Side view of one of the (b) $Ca_8C_{26}$ and (d) $Ca_6C_{32}$ relaxed configurations.

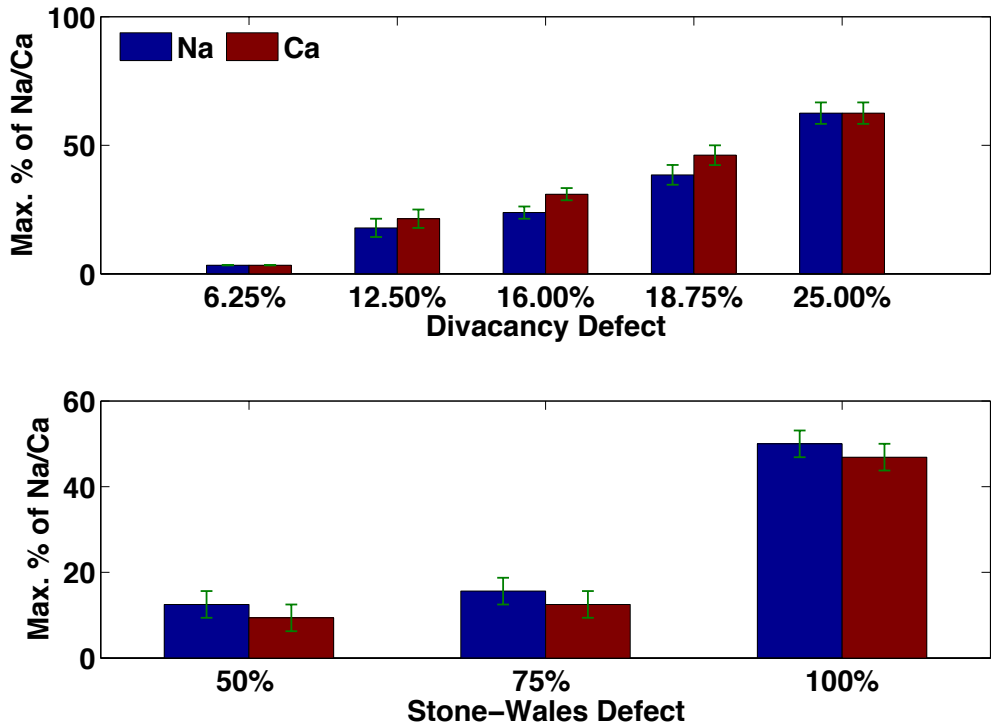

**Fig.7 Maximum percentages of Na/Ca adsorbed for different percentages of DV and SW defects.**

Fig.7 summarizes the maximum percentage of Na/Ca adsorbed for different percentages of DV and SW defect. Capacity $C$ (mAh/g) can be computed from percentage of adsorption $p$ as:

$$C = \frac{1}{A_c}\left[\left(\frac{p}{100}\right) \cdot v \cdot F \cdot 10^3\right] \qquad (5)$$

Where $p$ : Percentage of adsorption (in %)

$v$ : Vacancy (Na = 1; Ca = 2)

$F$ : Faraday Constant (26.801 Ah/Mole)

$A_c$ : Atomic mass of Carbon (=12.011)

For 6.25% DV defect, maximum percentage of adsorption is 6.67% corresponding to capacity of 148.8325 mAh/g for Na and 297.6649 mAh/g for Ca. With the increase in defect density, we obtain maximum percentage of adsorption for Na/Ca around 19%, 25%, 40% and 65% for 12.50%, 16%, 18.75% and 25% defects respectively. Hence for maximum defect density of 25%, we can obtain maximum capacity of around 1459 mAh/g for Na and 2900 mAh/g for Ca. For SW defect, maximum percentage of adsorption is around 10%, 13% and 48% for 50%, 75% and 100% SW defect respectively. Hence for 100% SW defect i.e. structure full of 5-7 rings, we can achieve maximum capacity of around 1071 mAh/g and 2142 mAh/g for Na and Ca respectively. We observe that for DV defect, capacity increases gradually with the increase of defect density. However, for SW defect, until the system reaches its maximum defect density i.e. system full of 5-7 rings, capacity does not increase much. This can be attributed to the fact that for Haeckelite structure, drop in equilibrium energy is drastic while for DV defects; drop in equilibrium energy is gradual.

**Conclusion**

In conclusion, we have performed the first-principles calculations to study the Na and Ca adsorption on graphene with various percentages of DV and SW defects. Our results show that adsorption is not possible in pristine graphene. But the presence of defects enhances the adsorption and the potential is more when the adatoms are on and around the defective zone. With the increase in defect density, maximum capacity obtained is much higher than that of graphite. This study will help create better anode materials which can replace graphite for higher capacity and better cycling performance for NIBs and CIBs.

1. Dahn, J. R.; Zheng, T.; Liu, Y. H.; Xue, J. S. *Science* **1995,** 270, (5236), 590-593.
2. Armand, M.; Tarascon, J. M. *Nature* **2008,** 451, (7179), 652-7.
3. Dunn, B.; Kamath, H.; Tarascon, J. M. *Science* **2011,** 334, (6058), 928-35.
4. J.M. Tarascon, M. A. *Nature* **2001,** 414, 359-367.
5. Tarascon, J. M.; Armand, M. *Nature* **2001,** 414, (6861), 359-67.
6. Teng, F. Z.; McDonough, W. F.; Rudnick, R. L.; Dalpé, C.; Tomascak, P. B.; Chappell, B. W.; Gao, S. *Geochimica et Cosmochimica Acta* **2004,** 68, (20), 4167-4178.
7. Palomares, V.; Serras, P.; Villaluenga, I.; Hueso, K. B.; Carretero-González, J.; Rojo, T. *Energy & Environmental Science* **2012,** 5, (3), 5884.

8.	Ong, S. P.; Chevrier, V. L.; Hautier, G.; Jain, A.; Moore, C.; Kim, S.; Ma, X. H.; Ceder, G. *Energy & Environmental Science* **2011,** 4, (9), 3680-3688.

9.	Kim, S.-W.; Seo, D.-H.; Ma, X.; Ceder, G.; Kang, K. *Advanced Energy Materials* **2012,** 2, (7), 710-721.

10.	Hayashi, M.; Arai, H.; Ohtsuka, H.; Sakurai, Y. *Journal of Power Sources* **2003,** 119, 617-620.

11.	Amatucci, G. G.; Badway, F.; Singhal, A.; Beaudoin, B.; Skandan, G.; Bowmer, T.; Plitza, I.; Pereira, N.; Chapman, T.; Jaworski, R. *Journal of The Electrochemical Society* **2001,** 148, (8), A940-A950.

12.	Lin, Y. M.; Abel, P. R.; Gupta, A.; Goodenough, J. B.; Heller, A.; Mullins, C. B. *ACS Appl Mater Interfaces* **2013**.

13.	Seyfried, W. E.; Janecky, D. R.; Mottl, M. J. *Geochimica et Cosmochimica Acta* **1984,** 48, (3), 557-569.

14.	Liu, J.; Zhang, J. G.; Yang, Z. G.; Lemmon, J. P.; Imhoff, C.; Graff, G. L.; Li, L. Y.; Hu, J. Z.; Wang, C. M.; Xiao, J.; Xia, G.; Viswanathan, V. V.; Baskaran, S.; Sprenkle, V.; Li, X. L.; Shao, Y. Y.; Schwenzer, B. *Advanced Functional Materials* **2013,** 23, (8), 929-946.

15.	Mortazavi, M.; Deng, J. K.; Shenoy, V. B.; Medhekar, N. V. *Journal of Power Sources* **2013,** 225, 207-214.

16.	Zhu, H.; Jia, Z.; Chen, Y.; Weadock, N.; Wan, J.; Vaaland, O.; Han, X.; Li, T.; Hu, L. *Nano Lett* **2013,** 13, 3093-3100.

17.	Sadoway, J. R. a. D. R. In *Towards the Development of Calcium-Ion Batteries*, PRiME 2012 , , Honolulu, 2012; The Electrochemical Society: Honolulu, 2012.

18.	Xu, J.; Lavan, D. A. *Nature Nanotechnology* **2008,** 3, (11), 666-670.

19.	Divincenzo, D. P.; Mele, E. J. *Physical Review B* **1985,** 32, (4), 2538-2553.

20.	Medeiros, P. V. C.; Mota, F. D.; Mascarenhas, A. J. S.; de Castilho, C. M. C. *Nanotechnology* **2010,** 21, (11), 115701.

21.	C. Ataca, E. A., S. Ciraci, H. Ustunel. *Applied Physics Letter* **2008,** 93.

22.	Stournara, M. E.; Shenoy, V. B. *Journal of Power Sources* **2011,** 196, (13), 5697-5703.

23. Shimoda, H.; Gao, B.; Tang, X. P.; Kleinhammes, A.; Fleming, L.; Wu, Y.; Zhou, O. *Physical Review Letters* **2002,** 88, (1).

24. Meunier, V.; Kephart, J.; Roland, C.; Bernholc, J. *Physical Review Letters* **2002,** 88, (7).

25. Chan, C. K.; Peng, H. L.; Liu, G.; McIlwrath, K.; Zhang, X. F.; Huggins, R. A.; Cui, Y. *Nature Nanotechnology* **2008,** 3, (1), 31-35.

26. Novoselov, K. S.; Geim, A. K.; Morozov, S. V.; Jiang, D.; Zhang, Y.; Dubonos, S. V.; Grigorieva, I. V.; Firsov, A. A. *Science* **2004,** 306, (5696), 666-669.

27. Lv, W.; Tang, D. M.; He, Y. B.; You, C. H.; Shi, Z. Q.; Chen, X. C.; Chen, C. M.; Hou, P. X.; Liu, C.; Yang, Q. H. *ACS Nano* **2009,** 3, (11), 3730-3736.

28. Yoo, E.; Kim, J.; Hosono, E.; Zhou, H.; Kudo, T.; Honma, I. *Nano Letters* **2008,** 8, (8), 2277-2282.

29. Jang, B. Z.; Liu, C. G.; Neff, D.; Yu, Z. N.; Wang, M. C.; Xiong, W.; Zhamu, A. *Nano Letters* **2011,** 11, (9), 3785-3791.

30. Florian Banhart, J. K., Arkady V. Krasheninnikov. *ACS Nano* **2011,** 5, (1), 26-41.

31. Hashimoto, A.; Suenaga, K.; Gloter, A.; Urita, K.; Iijima, S. *Nature* **2004,** 430, (7002), 870-873.

32. Terrones, H.; Lv, R.; Terrones, M.; Dresselhaus, M. S. *Rep Prog Phys* **2012,** 75, (6), 062501.

33. Meyer, J. C.; Kisielowski, C.; Erni, R.; Rossell, M. D.; Crommie, M. F.; Zettl, A. *Nano Letters* **2008,** 8, (11), 3582-3586.

34. Kotakoski, J.; Krasheninnikov, A. V.; Kaiser, U.; Meyer, J. C. *Physical Review Letters* **2011,** 106, (10), 105505.

35. Ugeda, M. M.; Brihuega, I.; Guinea, F.; Gomez-Rodriguez, J. M. *Physical Review Letters* **2010,** 104, (9), 096804.

36. Ugeda, M. M.; Brihuega, I.; Hiebel, F.; Mallet, P.; Veuillen, J. Y.; Gomez-Rodriguez, J. M.; Yndurain, F. *Physical Review B* **2012,** 85, (12).

37. Liu-Jiang Zhou, Z. F. H., Li-Ming Wu. *The Journal of Physical Chemistry C* **2012,** 116, 21780-21787.

38. Kresse, G.; Furthmuller, J. *Physical Review B* **1996,** 54, (16), 11169-11186.

39. Kresse, G.; Joubert, D. *Physical Review B* **1999,** 59, (3), 1758-1775.

40. Blochl, P. E. *Phys Rev B Condens Matter* **1994,** 50, (24), 17953-17979.

41. Perdew, J. P.; Burke, K.; Wang, Y. *Physical Review B* **1996,** 54, (23), 16533-16539.

42. Aydinol, M. K.; Ceder, G. *Journal of the Electrochemical Society* **1997,** 144, (11), 3832-3835.

43. Gass, M. H.; Bangert, U.; Bleloch, A. L.; Wang, P.; Nair, R. R.; Geim, A. K. *Nature Nanotechnology* **2008,** 3, (11), 676-681.

44. Krasheninnikov, A. V.; Lehtinen, P. O.; Foster, A. S.; Nieminen, R. M. *Chemical Physics Letters* **2006,** 418, (1-3), 132-136.

45. El-Barbary, A. A.; Telling, R. H.; Ewels, C. P.; Heggie, M. I.; Briddon, P. R. *Physical Review B* **2003,** 68, (14).

46. Lahiri, J.; Lin, Y.; Bozkurt, P.; Oleynik, I. I.; Batzill, M. *Nature Nanotechnology* **2010,** 5, (5), 326-329.

47. Carlsson, J. M.; Scheffler, M. *Physical Review Letters* **2006,** 96, (4), 046806.

48. Brunetto, G.; Autreto, P. A. S.; Machado, L. D.; Santos, B. I.; dos Santos, R. P. B.; Galvao, D. S. *Journal of Physical Chemistry C* **2012,** 116, (23), 12810-12813.

49. Li, L.; Reich, S.; Robertson, J. *Physical Review B* **2005,** 72, (18).

50. Terrones, H.; Terrones, M.; Hernandez, E.; Grobert, N.; Charlier, J. C.; Ajayan, P. M. *Phys Rev Lett* **2000,** 84, (8), 1716-9.

51. Fan, X. F.; Zheng, W. T.; Kuo, J. L. *ACS Applied Materials & Interfaces* **2012,** 4, (5), 2432-2438.

52. Yang, C.-K. *Applied Physics Letter* **2009,** 94.

53. Li, J.; Medhekar, N. V.; Shenoy, V. B. *The Journal of Physical Chemistry C* **2013,** 117, (30), 15842-15848.